\documentclass[%
 reprint, 
 amsmath,amssymb,
 aps,
]{revtex4-1}

\usepackage{graphicx}
\usepackage{dcolumn}
\usepackage{bm}
\usepackage{xcolor}

\usepackage{appendix}
\usepackage{placeins}

\begin{document}

\title{A Magnetic-like Model for Chemotactic Navigation in Ants 
}

\author{Rosa Flaquer-Galm\' es, Daniel Campos and Javier Crist\' in}

\affiliation{
 Grup de F\' isica Estad\' istica, Departament de F\' isica. Facultat de Ci\`encies, Universitat Aut\`onoma de Barcelona, 08193 Bellaterra (Barcelona), Spain }

\date{\today}

\begin{abstract}

We propose a physical framework for ant navigation of chemical trails. For this, we use controlled experiments in which individuals follow narrow pheromone trails, for which ants display oscillatory motion, as previously reported in the literature. We model this behavior by treating chemotaxis as an effective magnetic interaction between the ant velocity and the local chemical gradient. Under suitable approximations, the model yields an analytical expression for the velocity correlations in the direction perpendicular to the trail, predicting an underdamped oscillatory decay. This theoretical prediction is in qualitative agreement with our experimental measurements, indicating that the model captures the essential dynamical features of ant trail following. We fit the model parameters to individual trajectories in order to assess the consistency of the underlying assumptions, finding the same parameter relationship in both theory and experiment. Our results contribute to the characterization of chemotactic navigation in ants and illustrate how physical modeling can provide mechanistic insights into complex biological dynamics.
 \end{abstract}

\maketitle


\twocolumngrid
\section{Introduction} \label{sec:intro}

Living organisms navigate their environment through the continuous processing of external cues. Such strategies include the use of visual signals \cite{wehner1996visual, collett2002memory,seelig2015neural}, geomagnetic orientation in migratory birds \cite{cochran2004migrating,lohmann2004geomagnetic}, and echolocation in bats \cite{jensen2005echolocating,yovel2011complex}. 
Among these, chemotaxis (the ability to detect and respond to chemical gradients) is widespread across biological scales. Bacteria employ temporal and spatial comparisons of chemical concentrations to locate nutrients \cite{wadhams2004making, keegstra2022ecological}. Eukaryotic cells migrate along concentration gradients during processes such as the immune response \cite{swaney2010eukaryotic}. Multicellular organisms rely on chemical cues for reproduction or foraging, such as in mammalian sperm \cite{cohen1994sequential} or insect olfaction \cite{kadakia2022odour, reddy2022olfactory, emonet2024olfactory,renou2020insect}.


Ants are a paradigmatic example of chemotactic behavior, exhibiting a repertoire of responses to chemical cues at both the individual and collective levels \cite{attygalle1985ant, vander1988isolation}. Many species deposit pheromones along foraging routes, enabling either the same individual or others to follow trails during subsequent excursions or to navigate back to the nest \cite{lenoir2011trail}. Within the nest, chemical signals also guide spatial organization, with distinct “road signs” directing movement and ensuring efficient navigation even in the absence of visual landmarks or in complex environments \cite{heyman2017ants}. The translation of these local chemical cues into ant movement has motivated a variety of mathematical approaches. Some models focus on individual-level responses \cite{liebchen2018synthetic, riman2021dynamics}, while others describe collective dynamics through partial differential equations that capture the spatiotemporal distribution of ant groups \cite{keller1971model, alt1980biased, calenbuhr1992model, amorim2015modeling,ramakrishnan2014spatiotemporal}.


The notion of navigation along chemical gradients resonates with a physicist’s perspective. In mechanistic descriptions of motion, forces emerge from gradients in energetic landscapes, driving systems toward states of lower energy. This observation naturally suggests a mapping onto physical models, in which chemical gradients can be interpreted through effective forces. The idea of relating information processing to physical forces is not new, and similar mappings have been successfully applied in other biological contexts. In particular, collective alignment models treat interactions analogously to magnetic forces. In such models, agents (representing individual organisms) adjust their orientation in response to local fields, representing local interactions or external cues, capturing the emergence of large-scale patterns. Applications span across across  bird flocks \cite{bialek2012statistical, Attanasi2014information, mora2016local,cavagna2025spin}, fish schools \cite{katz2011inferring, mugica2022scale}, insect swarms \cite{cavagna2023natural}, bacterial colonies \cite{you2018geometry,jayathilake2017mechanistic}, to human crowds \cite{moussaid2011simple,cristin2019general}. Despite their widespread use, alignment-based models have not yet been applied to chemotaxis as an effective interaction with an external field.

In this work, we investigate individual ant navigation through this lens, characterizing chemotaxis as an effective force biasing movement along chemical gradients. To this end, we combine experimental observations of individual ants with a physical description of their motion, aiming to connect behavioral patterns with mechanistic principles. We first conduct a simple experiment to isolate essential features of ant movement in response to pheromone gradients, and as our main contribution, we present a physical model in which the driving forces arise from chemical gradients. The model, based on the previous Inertial Spin model (ISM) \cite{cavagna2015flocking}, captures the key features of the experimental data, shedding light on the underlying mechanisms. 

The article is organized as follows: Section~\ref{sec:exp} describes the experimental setup and analyzes the ant trajectories, reporting the oscillatory behavior along the trail. Section~\ref{sec:model} introduces the physical model and derives an analytical expression for the velocity correlations. Section~\ref{sec:res} compares these analytical predictions with the experimental results, showing good agreement between both, and discusses the broader implications of this modeling approach. Finally, Section~\ref{sec:disc} summarizes our work and outlines directions for future research.

\begin{figure*}[t!]
    \centering
    \includegraphics[width=\linewidth]{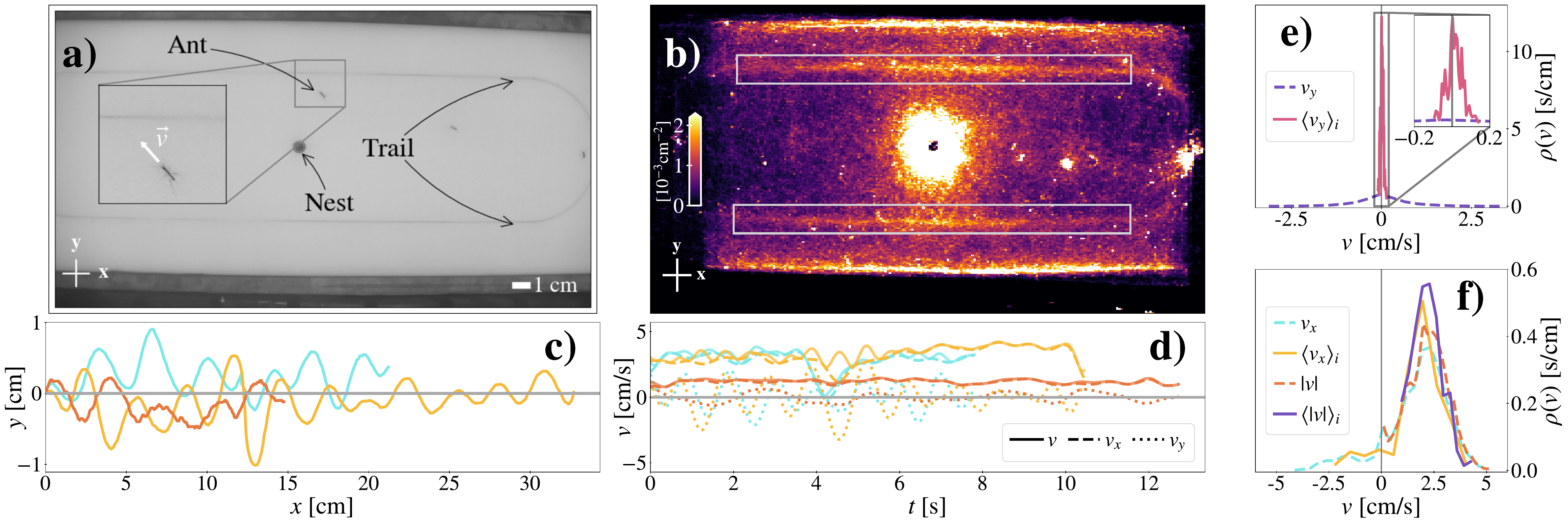}
    \caption{a) Experimental plate in the frame of a given experiment. b) Heatmap of ant occupancy in the arena, previous to data processing, including data from the $20$ daily experiments described in the main text. c) Trajectories of three different ants in the region near the pheromone trail, highlighted in a white rectangle in panel b). The horizontal gray line indicates the position of the center of the pheromone trail. d) Velocity signal of the three trajectories plotted in panel c), showing separately the velocity components along the trail direction ($v_x$) in dashed lines and the perpendicular direction ($v_y$) in dotted lines. In solid lines, we plot the total speed $v=\vert \vec{v} \vert$. e) Velocity component $v_y$ distribution (dashed purple line) after integrating all $156$ trajectories and individual mean velocity $\left< v_y\right>_i$ distribution (solid pink line). The inset shows a zoom of the central region. f) Velocity component $v_x$ (dashed light-blue line) and speed $v$ (dashed orange line) distribution after integrating all $156$ trajectories and individual mean velocity $\left< v_x\right>_i$ (yellow solid line)  and individual mean speed $\left< v \right>$ (purple solid line) distributions. }
    \label{fig:ex_traj}
\end{figure*}

\section{Experimental data} \label{sec:exp}

To obtain data of ants following a chemical trail, we used a simple design (Figure \ref{fig:ex_traj}a)). It consists of a plate connected to a nest of \textit{Aphaenogaster Senilis} ants (labeled \textit{Nest} in the figure). On the plate, we manually introduced a continuous pheromone trail, (labeled \textit{Trail}). The trail was designed to approximate an infinite straight line using two parallel segments joined by semicircles, forming an oval-shaped loop that effectively eliminates finite-trail effects by creating a virtually infinite path. To encourage ant exploration during successive experimental trials, a food patch was placed one of the curves of the trail. We have recorded one hour of ant activity in this setup over 20 days. Using the software AnTracks \cite{AnTracks}, we have obtained the trajectories of each ant present on the arena at a given moment. In Appendix \ref{app:setup} and \ref{app:tracking} we present in greater detail the  experimental setup and the obtention of the data trajectories.

From the trajectories of the ants during their exploration, we can analyze how they occupy the plate. The heatmap of ant occupancy (Figure \ref{fig:ex_traj}b)) shows the time spent in each region, averaged over all experiments. Four regions of higher density are clearly visible: the hole connecting the plate and the nest, the food patch, the plate borders, and the pheromone trail. The first is a trivial consequence of the setup configuration, as every ant that enters or exits the structure must do so through the hole. The second concerns the time spent gathering the food. The third is also expected, as it has been reported that animals in enclosed arenas tend to move toward the edges and borders to use the physical contact with them as a cognitive reference, an effect called thigmotaxis \cite{dussutour2005amplification,jeanson2003model}. The fourth and final region is the most relevant for our study: the pheromone trail is highlighted by the higher occupancy of the ants in that area. This suggests that the ants move, to some extent, preferentially towards regions of higher concentration of the chemical signal.

To characterize the behavior of the ants when navigating a pheromone trail, we focus on the dynamics close to the region defined by the boxed areas in Figure \ref{fig:ex_traj}b). The discussion about the specific definition of this region can be found in Appendix \ref{app:processing}. 
Small variations in this definition do not significantly affect the data.
In Figure \ref{fig:ex_traj}c) we show partial trajectories of three different ants in that region, to illustrate how the ants move along the trail for an extended period (videos of these trajectories can be found in the supplementary material online). We note that, while the ants move along the trail on average, they consistently oscillate from one side to the other. This behavior becomes more evident when analyzing the ants' velocities (see Fig. \ref{fig:ex_traj}d)), where we can see that the velocity component in the direction of the trail ($v_x$) is the main contributor to the ants' displacement. From the perpendicular component ($v_y$), in addition to being much smaller than $v_x$, it oscillates around $0$.

This feature is not a particular one of the trajectories shown, but is instead a property observed in the majority of them, suggesting that it is an evolutionarily orchestrated response of ants when navigating chemical trails \cite{Catania2013,Jayakumar2022,Popp2023}. In Figure \ref{fig:ex_traj}e), we present the distributions of velocity $v_y$, grouping all trajectories, and the individual mean velocity $\left< v_y\right>_i$, where $\left< f\right>_i = \frac{1}{m_i}\sum_j f_i^j$, where $m_i$ is the number of points in trajectory $i$. We observe that both distributions are symmetric and with a mean of $0$ [cm/s], but the variance of $v_y$ distribution is much larger.  In $\left< v_y\right>_i$, temporal fluctuations of each trajectory have been integrated, but not inter-individual variations. Then, the differences between distributions originate from temporal variability within each trajectory, rather than from the inter-individual heterogeneity. This is in perfect agreement with the presence of oscillations around the trail. 
The same analysis for the distribution of velocity $v_x$ (see Figure \ref{fig:ex_traj}f)) confirms that the net movement is along the direction of the trail, as its mean is $\left<v_x\right>=1.7 \pm 1.5$ [cm/s]. The comparison of $v_x$ and $\left< v_x \right>_i$ distributions indicates that the velocity $v_x$  does not exhibit significant temporal fluctuations, behaving mostly as a constant in the motion of the ant. This velocity $v_x$, however, varies considerably across the population, highlighting substantial inter-individual differences. This is in agreement with the experimental signals shown in Figure \ref{fig:ex_traj}d). Finally, a comparison between the distribution of $v_x$ and the speed $v=\vert \vec{v} \vert$ yields that the total velocity is dominated by $v_x$ (see Figure \ref{fig:ex_traj}f)), enabling the ant to move along the trail.

To summarize, we have observed that ants are capable of detecting and following the chemical signal of the pheromone trail oscillating around it  while moving at an almost constant speed. This result is consistent with previous reports of zigzag motion in ants and other insects \cite{collett2014scene,Namiki2016,Wechsler2023}, often linked to bilateral sensing \cite{Draft2018,Louis2024} and supposed to facilitate signal recovery or information gathering during navigation \cite{lent2013phase,Carde2021,Popp2023}.

\section{Physical Model} \label{sec:model}
Let us discuss the previous empirical results from a physical perspective. The oscillatory behavior observed while following the trail cannot be attributed to an external agent introducing a periodic signal, but is a characteristic strategy that ants use as a part of their navigation behavior. The observation that ants maintain a relatively constant speed throughout their trajectories suggests that zigzagging in ants is almost exclusively conducted through turning or reorientation, without significant changes in the propulsion force exerted by the organism. 
We aim to explore whether chemotaxis can be incorporated into a simple physical model while still capturing these key features observed in the experiments. Specifically, our goal is to understand the underlying mechanisms present in the system and how they guide the movement of the ants. In a mechanistic approach, these interactions can be represented as forces that govern the ants' trajectories. 

\subsection{The Inertial Spin model} \label{sec:ISM}
Mechanistic approaches have been successfully applied to the study of other biological systems, such as starling flocks \cite{bialek2012statistical,Attanasi2014information,mora2016local,cavagna2025spin}. In that system, individuals move at nearly constant speeds, and their velocity reorientations appear to be driven by alignment interactions with their neighbors. While the interactions in bird flocks differ significantly from the interactions of individual ants with a chemical signal, the fundamental principle remains similar.  Birds orient themselves to match the orientation of their nearest neighbors. Similarly, ants orient their movement based on the interaction with the chemical signal. Both phenomena can be understood as the interaction between an agent and an effective field: in the case of birds, the field is generated by the combined orientation of the neighbors, while in the case of ants, it is given by the chemical signal. This renders a common interpretation in terms of spin systems.

For bird flocks, the Inertial Spin model (ISM) \cite{cavagna2015flocking,cavagna2024DLT} provides a convenient physical description of the system dynamics. In that model, the individual is characterized by a constant modulus velocity $\vec{v}$ and a spin $\vec{s}$, which acts as the generator of velocity reorientations. The equations of the dynamics of the ISM for a given individual read 

\begin{align}
\frac{d\vec{r}}{dt} &= \vec{v} \label{eq:dr_ISM}\\ 
\frac{d\vec{v}}{dt} &= \frac{1}{\chi} \vec{s} \times \vec{v} \label{eq:dv_ISM}\\ 
\frac{d\vec{s}}{dt} &= \vec{v} \times \left(-\frac{dH}{d\vec{v}}\right) - \frac{\eta}{\chi} \vec{s} + \vec{\xi} \label{eq:ds_ISM}
\end{align},
where the Hamiltonian $H$ accounts for the interactions present in the system
. The velocity derivative, with the cross product, strictly conserves the modulus of the velocity ($v_0$), allowing only for changes in its orientation. The parameter $\chi$ is defined as the inertia. The dissipative term $-\eta/\chi \vec{s}$ ensures that, in the absence of interactions, the ant avoids closed loops. The white noise $\vec{\xi}$ introduces statistical fluctuations and is defined by the correlator 
\begin{equation*}
    \left< \vec{\xi}(t) \vec{\xi}(t_0)  \right> = 2d\eta T \delta(t-t'),\label{eq:noise}
\end{equation*} for a $d$-dimensional space. The ISM  has proven to be successful in reproducing different complex biological data, particularly in information propagation within starling flocks (a phenomenon closely linked to velocity reorientations) and the dynamical scaling of insect swarms \cite{cavagna2015silent,cavagna2024DLT,cavagna2023natural}.

Given the success of the ISM and the conceptual analogies between bird flocks and ants in a pheromone trail, applying this framework to our system seems to be well grounded. In a flock, individuals are assumed to align with the local field. In the magnetism context, this is called a ferromagnetic-like interaction. In our context, where chemical paths are being used as a navigation guide by the ants, it is reasonable to assume a ferromagnetic-like interaction between velocity and concentration gradient $\vec{\nabla}c$, as this would naturally lead the ant to move toward regions of higher concentration. This is supported by our experimental observations. In Figure \ref{fig:ex_traj}b)), we can see how the ants spend a significant amount of time in areas of higher concentration. Moreover, ants tend to follow pheromone trails once they encounter them. This can be interpreted as a mechanism that forces the ants’ velocity to be perpendicular to the concentration gradient upon reaching the maximum concentration. 
A ferromagnetic-like interaction alone cannot account for velocity being orthogonal to the field. In magnetic systems, however, certain interactions favor spin orientations perpendicular—rather than parallel or antiparallel—to the local field. The simplest example is the Dzyaloshinskii–Moriya (DM) interaction \cite{moriya1960anisotropic,camley2023consequences}, which introduces a cross-product coupling between spins that promotes non-collinear configurations. 
We assume that both interactions (ferromagnetic-like and DM-like) are present in our system, as their combination would not only guide the ant towards the pheromone trail but also ensure that it follows it once located. Then, we propose a Hamiltonian of the form 
\begin{equation}
    H= -J \vec{v} \cdot \vec{\nabla}c +\vec{D} \cdot \left( \vec{v} \times \vec{\nabla} c\right) , \label{eq:h_ants}
\end{equation}
where $J$ represents the strength of the alignment (or ferromagnetic) interaction, $\vec{D}$ is the DM-like vector, defined as $\vec{D}=D \hat{n}$, where $\hat{n}$  is the unit vector defining the rotation axis \cite{cheong2007multiferroics}. By substituting the Hamiltonian \eqref{eq:h_ants} into equation \eqref{eq:ds_ISM}, we have the set of equations of the ISM for the ant navigating a chemical signal landscape.

\subsection{Near-trail approximation} \label{sec:near}

Equations (\ref{eq:dr_ISM}-\ref{eq:ds_ISM}) in general cannot be solved analytically. The common procedure is to study some limiting cases where simplifying assumptions can yield analytical results. In this spirit, we note that concentration gradients in our experimental setup must primarily occur in the direction perpendicular to the trail (see Figure \ref{fig:ex_traj}a)), assuming we neglect small fluctuations in pheromone deposition. The experimental analysis showed that the ants navigate the trail (see Figure \ref{fig:ex_traj}b)). In this regime, we expect the dominant interaction to be the one maintaining the velocity perpendicular to the gradient (DM). This leads us to assume $D \gg J$, allowing us to study the case where only the DM interaction is present as first approximation. 
The rotation axis $\hat{n}$ is defined analogously as in \cite{cheong2007multiferroics}, $\hat{n}=\hat{l} \times \hat{d}$,  where $\hat{d}=\vec{d}/\vert\vec{d}\vert$ is the unitary vector in the direction of the shortest distance from the trail to the ant and  $\hat{l}=\vec{l}/\vert\vec{l}\vert$ is the director vector of the pheromone trail at the closest point of the trail to the ant. The vectors $\hat{d}$ and $\hat{l}$ are orthogonal.
In our experimental setup, the ants live in the xy-plane. Therefore, $\hat{l}=\hat{e}_x$, and $\hat{d}=y/\vert y \vert \hat{e}_y$, leading to $\hat{n}=-\hat{e}_z$ for $y < 0$ and $\hat{n}=\hat{e}_z$ for $y > 0$. To further proceed, we need to specify the shape of the pheromone signal. 
Near the trail, to first order of approximation, the pheromone profile can be assumed to vary linearly and towards the center yielding a symmetric triangular function with the peak situated at the center of the trail. This leads to a gradient of the form
\begin{equation}
   \vec{\nabla}c=p(y) \hat{e}_y, 
\end{equation} 
where $p(y)=p$ for $y < 0$, $p(y)=-p$ for $y > 0$ and $p(0)=0$. Under these conditions, the Hamiltonian \eqref{eq:h_ants} takes the form
\begin{equation}
    H \approx - D v_x p. \label{eq:h_approx}
\end{equation}

 We note that the velocity $\vec{v}$ has only $x$ and $y$ components, implying that the generator of its rotations, the spin $\vec{s}$, is restricted to have only a $z$-component. From this point onward, we treat the spin as a scalar $s$, without losing generality. We substitute Eq. \eqref{eq:h_approx} into Eq. \eqref{eq:ds_ISM} and we obtain the following component-wise equations:

\begin{align}
\frac{dv_x}{dt} &=-\frac{1}{\chi} s v_y \label{eq:dvx} \\
\frac{dv_y}{dt}&=\frac{1}{\chi} s v_x \label{eq:dvy} \\
\frac{ds}{dt} &= - v_yDp - \frac{\eta}{\chi} s + \xi. \label{eq:ds}
\end{align}

Rewriting the velocity components in terms of the angle $\theta$, defined as the angle between the $x$-axis and $\vec{v}$, we have $v_x=v_0 \cos \theta$ and $v_y=v_0 \sin \theta$, where $v_0$ is the constant speed. By taking this into account in Eq. \eqref{eq:dvx}, we obtain
\begin{equation}
    \frac{d \theta}{dt}= \frac{s}{\chi}. \label{eq:dtheta}
\end{equation}

Differentiating Eq. \eqref{eq:dtheta} a second time and plugging it into Eq. \eqref{eq:ds}, we obtain a closed equation for the evolution of the angle $\theta$ that reads
\begin{equation}
    \frac{d^2\theta}{dt^2}+\frac{\eta}{\chi}\frac{d \theta}{dt}+ \frac{Dpv_0}{\chi} \sin \theta=\xi^*, \label{eq:theta}
\end{equation}
where $\xi^*=\xi/\chi$.

In section \ref{sec:exp} we have discussed how the speed is dominated by the $x$ component. Therefore, the angle $\theta$ is small, so we can write $\sin \theta \approx \theta$. Rewriting Eq. \eqref{eq:theta} accordingly yields

\begin{equation}
    \frac{d^2\theta}{dt^2}+\frac{\eta}{\chi}\frac{d \theta}{dt}+\frac{Dpv_0}{\chi}  \theta=\xi^*. \label{eq:small_theta}
\end{equation}

We note that Equation \eqref{eq:small_theta} corresponds to a stochastic damped harmonic oscillator. By defining

\begin{align}
\gamma &= \frac{\eta}{2\chi} \label{eq:gamma}\\ 
\omega_0^2 &= \frac{Dpv_0}{\chi}-\left(\frac{\eta}{2\chi}\right)^2,\label{eq:omega}
\end{align} 

one finds that the solution of \eqref{eq:small_theta} takes the form

\begin{equation}
    \theta(t) = \theta_h(t) + \int_0^tg(t,\tau)\xi^*(\tau)d\tau,
    \label{eq:green}
\end{equation}
where $g(t,\tau) = e^{-\gamma(t-\tau)}\sin\left( w_0\left(t - \tau \right) \right)\Theta(t-\tau)/w_0$ is the Green function of the damped harmonic oscillator and $\theta_h(t) = \theta_0e^{-\gamma t}\left( \cos(\omega_0t) + \gamma/\omega_0\sin(\omega_0t) \right)$ is the time evolution of the damped harmonic oscillator without noise. By considering the stochastic term, the integrated temporal behavior is encoded in the temporal correlations of the angle $\theta$:
\begin{equation}
  C_\theta(t)=  \frac{\left< \theta(0) \theta(t) \right>}{\left< \theta(0)^2 \right>}=e^{-\gamma t} \left( cos \left(\omega_0 t\right) +\frac{\gamma}{\omega_0} sin \left(\omega_0 t\right)\right),\label{eq:corr}
\end{equation}

We note that $v_y=v_0 \sin \theta\approx v_0 \theta$. Therefore, velocity correlations satisfy
\begin{equation}
C_\theta(t) \approx C_{v_y}(t).
\end{equation}

From equation \eqref{eq:corr} we observe that the velocity correlations have an oscillatory behavior with an envelope that decays exponentially in time. This exponential envelope is characterized by the damping parameter $\gamma$, while the oscillatory frequency is characterized by $\omega_0$. 

\begin{figure*}[t!]
    \centering
    \includegraphics[width=\linewidth]{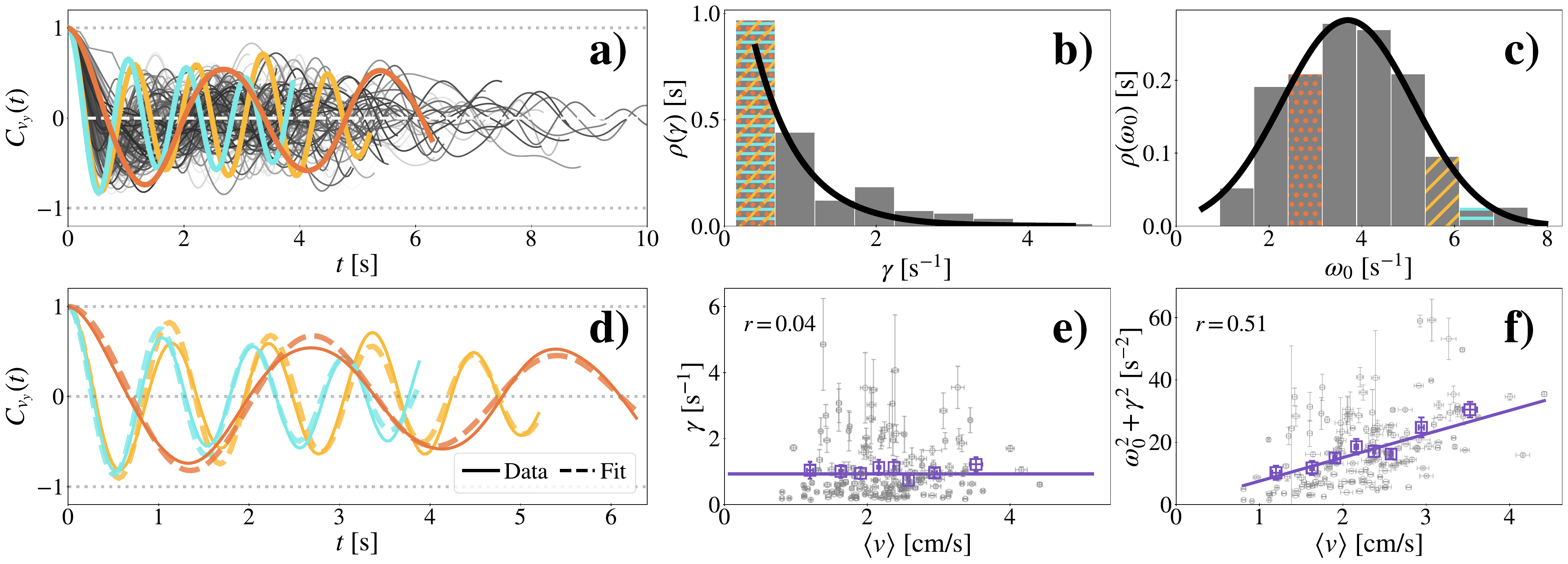}
    \caption{a) Velocity correlations in the y-direction (perpendicular to the trail) for the $156$ experimental trajectories. Lines are shown up to half the duration of each trajectory. The three experiments shown in Figure \ref{fig:ex_traj} have been highlighted in light-blue, orange, and yellow colors. b) and c) Probability density distribution of the obtained parameters $\gamma$ (b) and $\omega_0$ (c) from individually fitting each of the $156$ trajectories. The colored bars highlight the parameter values corresponding to the fits in d). d) Velocity correlations of the highlighted experiments (in solid lines) and the best fit for each of them according to equation \eqref{eq:corr} (in dashed lines). e) and f) The grey points represent the fitted values of $\gamma$ (e), and $\omega_0^2 +\gamma^2$ (f) as a function of the characteristic mean speed $\left<v\right>$ of each trajectory. The purple squared points correspond to an average of $20$ points, grouping them according to their $\left< v \right>$ value. The purple line corresponds to the best fit of the purple squared points to $\gamma=\eta/2\chi$ in (e) and to $\omega_0^2+\gamma^2=Dpv_0/\chi$ in (f), where $\left<v\right>$ takes the role of $v_0$. The value $r$ corresponds to the correlation coefficient of the gray points.}
    \label{fig:triple}
\end{figure*}
We note that the presence of oscillatory behavior depends on the model parameters. If $Dpv_0/\chi < \eta/ \left(2\chi\right)^2$, $\omega_0$ is imaginary and the expression in \eqref{eq:corr} becomes exponential. The overdamped version of equation \eqref{eq:ds_ISM}, where the inertial term is neglected with respect to the damping one, also yields an exponential decay without oscillations, with the form (see Appendix \ref{app:overdamped}).
\begin{equation}
C_{v_y}(t)=e^{-\gamma v_0 t}.\label{eq:corr_ov}
\end{equation}

To summarize, we have obtained an analytical expression for the correlations of the velocity $v_y$, adapting the ISM to ants navigating a chemical landscape, assuming they are near the trail. The correlations show an oscillatory behavior with an exponential envelope if the inertial term is not negligible and the dissipation does not dominate.

\section{Matching Theory and Experiments} \label{sec:res}

The analysis in Section \ref{sec:near}, eventually leading to Eq. \eqref{eq:corr}, provides an insight into the mechanisms that can generate oscillations within the ISM framework. More importantly, it explicitly predicts how these oscillations should manifest in the temporal correlations of velocity in the direction perpendicular to the trail (y-direction). Our experiments allow an analysis of this kind, as we have access to the time series of velocity (see Figure \ref{fig:ex_traj}d)). Our set of experimental trajectories provides a biological dataset where the ISM analytical predictions can be tested. 

In Figure \ref{fig:triple}a), we show the experimental velocity correlations $C_{v_y}(t)$ for all trajectories near the trail. We observe that most exhibit a temporal decay to zero with a distinct oscillatory pattern, with notable variability in the oscillation frequency and the decay rate. In color, we highlight the same three trajectories as in Figures \ref{fig:ex_traj}b) and \ref{fig:ex_traj}c), to help visualize this particular behavior.

We proceed to fit the experimental correlations with Equation \eqref{eq:corr}. We obtain a good agreement, with the $156$ trajectories having a nonlinear regression coefficient $R_a^2 \ge 0.4$, with an average coefficient $\left<R_a^2\right>=0.75$ and a median $R_{a,M}^2=0.76$ (see Appendix \ref{app:fit} for more detailed discussion of the fitting process and its results). The distribution of the fitted parameters $\gamma$ and $\omega_0$ is shown in Figures \ref{fig:triple}b) and \ref{fig:triple}c). The fitted parameters for the highlighted trajectories in Figure \ref{fig:triple}a) are presented in a different color of the histogram bars, while a visual comparison of the fit and the correlation function is shown in Figure \ref{fig:triple}d). We observe a manifest variability of these parameters within the ant population. 

The distribution of the damping parameter $\gamma$ yields an exponential-like shape, with a decay characterized by $\gamma_c=0.6 \pm 0.1$ [s$^{-1}$]. This suggests that high damping values, which would inhibit oscillations, are unlikely in ant dynamics. On the other hand, the distribution of the oscillation frequency $\omega_0$ resembles a Gaussian, centered around an average frequency of $\omega_0^c=3.70 \pm 0.01$ [s$^{-1}$]. The characteristic oscillation frequency might be related to the ants’ cognitive mechanisms, the constraints of the experimental setup, and individual variability  giving rise to the dispersion around the optimum. While the origin of this specific frequency remains an open question, oscillatory behavior has been proposed to enhance navigational parallax \cite{lent2013phase} or improved signal recovery following loss \cite{Carde2021,Popp2023}, especially in contexts combining visual and chemotactic cues \cite{zeil2014looking,Jayakumar2022,clement2023intrinsic,emonet2024olfactory}.

In Section \ref{sec:near}, we made several hypotheses that lead to \eqref{eq:corr}, which predicts oscillations in $C_{v_{y}}(t)$ for a certain regime of the $\gamma$ and $\omega_0$ parameters (underdamped regime), while parameters outside of this regime predict an exponential decay of $C_{v_{y}}(t)$ (overdamped regime). Our fitting analysis suggests that our experimental dataset is compatible with the oscillatory regime.  The values of the parameters $\gamma$ and $\omega_0$ will ultimately depend on the values of interactions, speed, dissipation, and inertia. Different Hamiltonian proposals, along with different assumptions, may also be able to predict the oscillatory behavior encoded in Eq. \eqref{eq:corr}, but with a different version of Eqs. \eqref{eq:gamma} and \eqref{eq:omega}. Given that, we seek to verify the validity of our model through comparison with the experimental results. For this, note that the values of $D$, $p$, $\eta$, and $\chi$ cannot be directly compared with empirical data; still, the constant speed $v_0$ can be related to the average experimental speed $\left< v \right>$ of each trajectory. Accordingly, we can test that the parameter $\gamma=\eta / 2 \chi$ is independent of speed; empirical data seems to confirm this result (Figure \ref{fig:triple}e)). For the case of $\omega_0$, we can reformulate Eqs. (\eqref{eq:gamma} - \eqref{eq:omega}) as $\omega_0^2+\gamma^2=Dpv_0/\chi$. Figure \ref{fig:triple}f) demonstrates that the corresponding linear dependence on speed is consistent with experiments. Altogether, the analysis suggests that our assumptions are reasonable and accurately capture the experimental behavior observed. For completness, the values of $\omega_0$ vs $\left< v \right>$ are included in Appendix \ref{app:omega_vs_v}.

We have checked (see Appendix \ref{app:fit}) whether there is any dependence between the trajectory duration and the fitted parameters to discard the possibility that nonstationary effects in the system are distorting our conclusions. We find that trajectory duration does not significantly affect the results, so we discard this possibility. This confirms that (i) trajectories used are sufficiently long to eliminate strong finite-length effects, and (ii) the dependence of $\omega_0$ on the average experimental speed $\left< v \right>$ in our experiments is not an artifact of finite trajectories. 

\section{Conclusions} \label{sec:disc}

Our analytical predictions for the ISM model are fundamentally consistent with the experimental observations, particularly regarding the presence of velocity oscillations, their temporal evolution, and the relation between the oscillation frequency and the characteristic speed of the ants. These findings lead us to conclude that the ISM framework, with a magnetic-like description of chemotaxis (DM interaction), effectively captures the fundamental mechanisms underlying chemotactic ant dynamics during trail following. 

This work could serve as the foundation for several research directions. The interplay between the DM term and the ferromagnetic term, which may play a fundamental role in locating the trail, remains to be investigated. The biological significance of the physical parameters and their absolute values are yet to be fully explored, and a way to do so may be to examine whether the observed parameter variability depends on the experimental context or reflects adaptive behavior by ants during navigation. Another research direction may rely on the introduction of autochemotaxis in the framework, where the pheromone concentration $c(x,t)$ would depend on the previous trajectory of the ant. We note that the framework presented here is not limited to chemical signals and can be extended to navigational scenarios guided by alternative sensory cues.

Overall, we believe this work highlights that even nontrivial biological behaviors, such as chemotaxis, can be effectively described using relatively simple physical principles. This underscores the broader applicability of physics-based approaches to understanding biological phenomena.

\section*{Acknowledgments}
The Authors acknowledge the financial support of the Spanish government under grant PID2021-122893NB-C22. We thank C. Navau and M. Veca for fruitful discussions.

\section*{Data Availability}
The data that support the findings of this article are openly available in \cite{rosa_github}.

\bibliography{biblio}

@misc{AnTracks,
    title = {AnTracks freeware},
    author = {Martin Stumpe},
    howpublished = {https://sites.google.com/view/antracks/home?authuser=0},
}

@article{collett2014scene,
  title={Scene perception and the visual control of travel direction in navigating wood ants},
  author={Collett, Thomas S and Lent, David D and Graham, Paul},
  journal={Philosophical Transactions of the Royal Society B: Biological Sciences},
  volume={369},
  number={1636},
  pages={20130035},
  year={2014},
  publisher={The Royal Society}
}

@article{zeil2014looking,
  title={Looking and homing: how displaced ants decide where to go},
  author={Zeil, Jochen and Narendra, Ajay and St{\"u}rzl, Wolfgang},
  journal={Philosophical Transactions of the Royal Society B: Biological Sciences},
  volume={369},
  number={1636},
  pages={20130034},
  year={2014},
  publisher={The Royal Society}
}

@article{cavagna2023natural,
  title={Natural swarms in 3.99 dimensions},
  author={Cavagna, Andrea and Di Carlo, Luca and Giardina, Irene and Grigera, Tom{\'a}s S and Melillo, Stefania and Parisi, Leonardo and Pisegna, Giulia and Scandolo, Mattia},
  journal={Nature Physics},
  volume={19},
  number={7},
  pages={1043--1049},
  year={2023},
  publisher={Nature Publishing Group UK London}
}

@article{clement2023intrinsic,
  title={An intrinsic oscillator underlies visual navigation in ants},
  author={Clement, Leo and Schwarz, Sebastian and Wystrach, Antoine},
  journal={Current Biology},
  volume={33},
  number={3},
  pages={411--422},
  year={2023},
  publisher={Elsevier}
}

@article{lent2013phase,
  title={Phase-dependent visual control of the zigzag paths of navigating wood ants},
  author={Lent, David D and Graham, Paul and Collett, Thomas S},
  journal={Current Biology},
  volume={23},
  number={23},
  pages={2393--2399},
  year={2013},
  publisher={Elsevier}
}

@article{cavagna2015flocking,
  title={Flocking and turning: a new model for self-organized collective motion},
  author={Cavagna, Andrea and Del Castello, Lorenzo and Giardina, Irene and Grigera, Tomas and Jelic, Asja and Melillo, Stefania and Mora, Thierry and Parisi, Leonardo and Silvestri, Edmondo and Viale, Massimiliano and others},
  journal={Journal of Statistical Physics},
  volume={158},
  pages={601--627},
  year={2015},
  publisher={Springer}
}

@article{cavagna2015silent,
  title={Silent flocks: constraints on signal propagation across biological groups},
  author={Cavagna, Andrea and Giardina, Irene and Grigera, Tomas S and Jelic, Asja and Levine, Dov and Ramaswamy, Sriram and Viale, Massimiliano},
  journal={Physical Review Letters},
  volume={114},
  number={21},
  pages={218101},
  year={2015},
  publisher={APS}
}

@article{cavagna2024DLT,
  title={Discrete Laplacian thermostat for flocks and swarms: the fully conserved Inertial Spin Model},
  author={Cavagna, Andrea and Cristín, Javier and Giardina, Irene and  Grigera, Tomás S and Veca, Mario},
  journal={Journal of Physics A: Mathematical and Theoretical},
  volume={57},
  number={41},
  pages={},
  year={2024},
  publisher={IOP Science}
}

@article{bialek2012statistical,
  title={Statistical mechanics for natural flocks of birds},
  author={Bialek, William and Cavagna, Andrea and Giardina, Irene and Mora, Thierry and Silvestri, Edmondo and Viale, Massimiliano and Walczak, Aleksandra M},
  journal={Proceedings of the National Academy of Sciences},
  volume={109},
  number={13},
  pages={4786--4791},
  year={2012},
  publisher={National Academy of Sciences}
}

@article{jeanson2003model,
  title={A model of animal movements in a bounded space},
  author={Jeanson, Rapha{\"e}l and Blanco, St{\'e}phane and Fournier, Richard and Deneubourg, Jean-Louis and Fourcassi{\'e}, Vincent and Theraulaz, Guy},
  journal={Journal of Theoretical Biology},
  volume={225},
  number={4},
  pages={443--451},
  year={2003},
  publisher={Elsevier}
}

@article{dussutour2005amplification,
  title={Amplification of individual preferences in a social context: the case of wall-following in ants},
  author={Dussutour, Audrey and Deneubourg, Jean-Louis and Fourcassi{\'e}, Vincent},
  journal={Proceedings of the Royal Society B: Biological Sciences},
  volume={272},
  number={1564},
  pages={705--714},
  year={2005},
  publisher={The Royal Society London}
}

@article{mora2016local,
  title={Local equilibrium in bird flocks},
  author={Mora, Thierry and Walczak, Aleksandra M and Del Castello, Lorenzo and Ginelli, Francesco and Melillo, Stefania and Parisi, Leonardo and Viale, Massimiliano and Cavagna, Andrea and Giardina, Irene},
  journal={Nature physics},
  volume={12},
  number={12},
  pages={1153--1157},
  year={2016},
  publisher={Nature Publishing Group UK London}
}

@article{renou2020insect,
  title={Insect olfactory communication in a complex and changing world},
  author={Renou, Michel and Anton, Sylvia},
  journal={Current Opinion in Insect Science},
  volume={42},
  pages={1--7},
  year={2020},
  publisher={Elsevier}
}

@article{Attanasi2014information,
  title={Information transfer and behavioural inertia in starling flocks},
  author={Attanasi, Alessandro and Cavagna, Andrea and Del Castello, Lorenzo and Giardina, Irene and Grigera, Tomas S and Jeli{\'c}, Asja and Melillo, Stefania and Parisi, Leonardo and Pohl, Oliver and Shen, Edward and others},
  journal={Nature Physics},
  volume={10},
  number={9},
  pages={691--696},
  year={2014},
  publisher={Nature Publishing Group UK London}
}

@article{cavagna2025spin,
  title={Spin-Waves without Spin-Waves: A Case for Soliton Propagation in Starling Flocks},
  author={Cavagna, Andrea and Cimino, Guido and Crist{\'\i}n, Javier and Fiorini, Matteo and Giardina, Irene and Giustiniani, Angelo and Grigera, Tom{\'a}s S and Melillo, Stefania and Palombella, Roberto A and Parisi, Leonardo and others},
  journal={arXiv preprint arXiv:2505.19665},
  year={2025}
}

@article{ramakrishnan2014spatiotemporal,
  title={Spatiotemporal chemotactic model for ant foraging},
  author={Ramakrishnan, Subramanian and Laurent, Thomas and Kumar, Manish and Bertozzi, Andrea L},
  journal={Modern Physics Letters B},
  volume={28},
  number={30},
  pages={1450238},
  year={2014},
  publisher={World Scientific}
}

@article{cristin2019general,
  title={General scaling in bidirectional flows of self-avoiding agents},
  author={Crist{\'\i}n, Javier and Mendez, Vicenc and Campos, Daniel},
  journal={Scientific reports},
  volume={9},
  number={1},
  pages={18488},
  year={2019},
  publisher={Nature Publishing Group UK London}
}

@article{jayathilake2017mechanistic,
  title={A mechanistic Individual-based Model of microbial communities},
  author={Jayathilake, Pahala Gedara and Gupta, Prashant and Li, Bowen and Madsen, Curtis and Oyebamiji, Oluwole and Gonz{\'a}lez-Cabaleiro, Rebeca and Rushton, Steve and Bridgens, Ben and Swailes, David and Allen, Ben and others},
  journal={PloS one},
  volume={12},
  number={8},
  pages={e0181965},
  year={2017},
  publisher={Public Library of Science San Francisco, CA USA}
}

@article{you2018geometry,
  title={Geometry and mechanics of microdomains in growing bacterial colonies},
  author={You, Zhihong and Pearce, Daniel JG and Sengupta, Anupam and Giomi, Luca},
  journal={Physical Review X},
  volume={8},
  number={3},
  pages={031065},
  year={2018},
  publisher={APS}
}

@article{moussaid2011simple,
  title={How simple rules determine pedestrian behavior and crowd disasters},
  author={Moussa{\"\i}d, Mehdi and Helbing, Dirk and Theraulaz, Guy},
  journal={Proceedings of the National Academy of Sciences},
  volume={108},
  number={17},
  pages={6884--6888},
  year={2011},
  publisher={National Academy of Sciences}
}

@article{camley2023consequences,
  title={Consequences of the Dzyaloshinskii-Moriya interaction},
  author={Camley, Robert E and Livesey, Karen L},
  journal={Surface Science Reports},
  volume={78},
  number={3},
  pages={100605},
  year={2023},
  publisher={Elsevier}
}

@article{moriya1960anisotropic,
  title={Anisotropic superexchange interaction and weak ferromagnetism},
  author={Moriya, T{\^o}ru},
  journal={Physical Review},
  volume={120},
  number={1},
  pages={91},
  year={1960},
  publisher={APS}
}

@article{reddy2022olfactory,
  title={Olfactory sensing and navigation in turbulent environments},
  author={Reddy, Gautam and Murthy, Venkatesh N and Vergassola, Massimo},
  journal={Annual Review of Condensed Matter Physics},
  volume={13},
  number={1},
  pages={191--213},
  year={2022},
  publisher={Annual Reviews}
}

@article{emonet2024olfactory,
  title={Olfactory cues and memories in animal navigation},
  author={Emonet, Thierry and Vergassola, Massimo},
  journal={Nature Reviews Physics},
  volume={6},
  number={4},
  pages={215--216},
  year={2024},
  publisher={Nature Publishing Group UK London}
}

@article{kadakia2022odour,
  title={Odour motion sensing enhances navigation of complex plumes},
  author={Kadakia, Nirag and Demir, Mahmut and Michaelis, Brenden T and DeAngelis, Brian D and Reidenbach, Matthew A and Clark, Damon A and Emonet, Thierry},
  journal={Nature},
  volume={611},
  number={7937},
  pages={754--761},
  year={2022},
  publisher={Nature Publishing Group UK London}
}

@article{katz2011inferring,
  title={Inferring the structure and dynamics of interactions in schooling fish},
  author={Katz, Yael and Tunstr{\o}m, Kolbj{\o}rn and Ioannou, Christos C and Huepe, Cristi{\'a}n and Couzin, Iain D},
  journal={Proceedings of the National Academy of Sciences},
  volume={108},
  number={46},
  pages={18720--18725},
  year={2011},
  publisher={National Academy of Sciences}
}

@article{mugica2022scale,
  title={Scale-free behavioral cascades and effective leadership in schooling fish},
  author={M{\'u}gica, Julia and Torrents, Jordi and Crist{\'\i}n, Javier and Puy, Andreu and Miguel, M Carmen and Pastor-Satorras, Romualdo},
  journal={Scientific reports},
  volume={12},
  number={1},
  pages={10783},
  year={2022},
  publisher={Nature Publishing Group UK London}
}

@article{cheong2007multiferroics,
  title={Multiferroics: a magnetic twist for ferroelectricity},
  author={Cheong, Sang-Wook and Mostovoy, Maxim},
  journal={Nature Materials},
  volume={6},
  number={1},
  pages={13--20},
  year={2007},
  publisher={Nature Publishing Group UK London}
}

@article{wehner1996visual,
  title={Visual navigation in insects: coupling of egocentric and geocentric information},
  author={Wehner, R{\"u}diger and Michel, Barbara and Antonsen, Per},
  journal={Journal of Experimental Biology},
  volume={199},
  number={1},
  pages={129--140},
  year={1996},
  publisher={The Company of Biologists Ltd}
}

@article{collett2002memory,
  title={Memory use in insect visual navigation},
  author={Collett, Thomas S and Collett, Matthew},
  journal={Nature Reviews Neuroscience},
  volume={3},
  number={7},
  pages={542--552},
  year={2002},
  publisher={Nature Publishing Group UK London}
}

@article{seelig2015neural,
  title={Neural dynamics for landmark orientation and angular path integration},
  author={Seelig, Johannes D and Jayaraman, Vivek},
  journal={Nature},
  volume={521},
  number={7551},
  pages={186--191},
  year={2015},
  publisher={Nature Publishing Group UK London}
}

@article{cochran2004migrating,
  title={Migrating songbirds recalibrate their magnetic compass daily from twilight cues},
  author={Cochran, William W and Mouritsen, Henrik and Wikelski, Martin},
  journal={Science},
  volume={304},
  number={5669},
  pages={405--408},
  year={2004},
  publisher={American Association for the Advancement of Science}
}

@article{lohmann2004geomagnetic,
  title={Geomagnetic map used in sea-turtle navigation},
  author={Lohmann, Kenneth J and Lohmann, Catherine MF and Ehrhart, Llewellyn M and Bagley, Dean A and Swing, Timothy},
  journal={Nature},
  volume={428},
  number={6986},
  pages={909--910},
  year={2004},
  publisher={Nature Publishing Group UK London}
}

@article{jensen2005echolocating,
  title={Echolocating bats can use acoustic landmarks for spatial orientation},
  author={Jensen, Marianne Egebjerg and Moss, Cynthia F and Surlykke, Annemarie},
  journal={Journal of Experimental Biology},
  volume={208},
  number={23},
  pages={4399--4410},
  year={2005},
  publisher={Company of Biologists}
}

@article{yovel2011complex,
  title={Complex echo classification by echo-locating bats: a review},
  author={Yovel, Yossi and Franz, Matthias O and Stilz, Peter and Schnitzler, Hans-Ulrich},
  journal={Journal of Comparative Physiology A},
  volume={197},
  pages={475--490},
  year={2011},
  publisher={Springer}
}

@article{wadhams2004making,
  title={Making sense of it all: bacterial chemotaxis},
  author={Wadhams, George H and Armitage, Judith P},
  journal={Nature Reviews Molecular Cell Biology},
  volume={5},
  number={12},
  pages={1024--1037},
  year={2004},
  publisher={Nature Publishing Group UK London}
}

@article{keegstra2022ecological,
  title={The ecological roles of bacterial chemotaxis},
  author={Keegstra, Johannes M and Carrara, Francesco and Stocker, Roman},
  journal={Nature Reviews Microbiology},
  volume={20},
  number={8},
  pages={491--504},
  year={2022},
  publisher={Nature Publishing Group UK London}
}

@article{swaney2010eukaryotic,
  title={Eukaryotic chemotaxis: a network of signaling pathways controls motility, directional sensing, and polarity},
  author={Swaney, Kristen F and Huang, Chuan-Hsiang and Devreotes, Peter N},
  journal={Annual Review of Biophysics},
  volume={39},
  number={1},
  pages={265--289},
  year={2010},
  publisher={Annual Reviews}
}

@article{cohen1994sequential,
  title={Sequential acquisition of chemotactic responsiveness by human spermatozoa},
  author={Cohen-Dayag, Anat and Ralt, Dina and Tur-Kaspa, Ilan and Manor, Mira and Makler, Amnon and Dor, Jehoshua and Mashiach, Shlomo and Eisenbach, Michael},
  journal={Biology of Reproduction},
  volume={50},
  number={4},
  pages={786--790},
  year={1994},
  publisher={Oxford University Press}
}

@incollection{attygalle1985ant,
  title={Ant trail pheromones},
  author={Attygalle, Athula B and Morgan, E David},
  booktitle={Advances in Insect Physiology},
  volume={18},
  pages={1--30},
  year={1985},
  publisher={Elsevier}
}

@article{lenoir2011trail,
  title={Trail-following behaviour in two Aphaenogaster ants},
  author={Lenoir, Alain and Benoist, Am{\'e}lie and Hefetz, Abraham and Francke, Wittko and Cerd{\'a}, Xim and Boulay, Rapha{\"e}l},
  journal={Chemoecology},
  volume={21},
  pages={83--88},
  year={2011},
  publisher={Springer}
}

@article{vander1988isolation,
  title={Isolation of the trail recruitment pheromone of Solenopsis invicta},
  author={Vander Meer, Robert K and Alvarez, Francisco and Lofgren, Clifford S},
  journal={Journal of Chemical Ecology},
  volume={14},
  pages={825--838},
  year={1988},
  publisher={Springer}
}

@article{heyman2017ants,
  title={Ants regulate colony spatial organization using multiple chemical road-signs},
  author={Heyman, Yael and Shental, Noam and Brandis, Alexander and Hefetz, Abraham and Feinerman, Ofer},
  journal={Nature Communications},
  volume={8},
  number={1},
  pages={15414},
  year={2017},
  publisher={Nature Publishing Group UK London}
}

@article{keller1971model,
  title={Model for chemotaxis},
  author={Keller, Evelyn F and Segel, Lee A},
  journal={Journal of Theoretical Biology},
  volume={30},
  number={2},
  pages={225--234},
  year={1971},
  publisher={Elsevier}
}

@article{alt1980biased,
  title={Biased random walk models for chemotaxis and related diffusion approximations},
  author={Alt, Wolgang},
  journal={Journal of Mathematical Biology},
  volume={9},
  pages={147--177},
  year={1980},
  publisher={Springer}
}

@article{liebchen2018synthetic,
  title={Synthetic chemotaxis and collective behavior in active matter},
  author={Liebchen, Benno and Lowen, Hartmut},
  journal={Accounts of Chemical Research},
  volume={51},
  number={12},
  pages={2982--2990},
  year={2018},
  publisher={ACS Publications}
}

@article{calenbuhr1992model,
  title={A model for osmotropotactic orientation (I)},
  author={Calenbuhr, Volker and Deneubourg, J-L},
  journal={Journal of Theoretical Biology},
  volume={158},
  number={3},
  pages={359--393},
  year={1992},
  publisher={Elsevier}
}

@article{amorim2015modeling,
  title={Modeling ant foraging: A chemotaxis approach with pheromones and trail formation},
  author={Amorim, Paulo},
  journal={Journal of Theoretical Biology},
  volume={385},
  pages={160--173},
  year={2015},
  publisher={Elsevier}
}

@article{riman2021dynamics,
  title={The dynamics of bilateral olfactory search and navigation},
  author={Riman, Nour and Victor, Jonathan D and Boie, Sebastian D and Ermentrout, Bard},
  journal={SIAM Review},
  volume={63},
  number={1},
  pages={100--120},
  year={2021},
  publisher={SIAM}
}

@article{Namiki2016,
title = {The neurobiological basis of orientation in insects: insights from the silkmoth mating dance},
journal = {Current Opinion in Insect Science},
volume = {15},
pages = {16-26},
year = {2016},
note = {Pests and resistance * Behavioural ecology},
issn = {2214-5745},
doi = {https://doi.org/10.1016/j.cois.2016.02.009},
url = {https://www.sciencedirect.com/science/article/pii/S2214574516300116},
author = {Shigehiro Namiki and Ryohei Kanzaki}
}

@article{Carde2021,
   author = "Cardé, Ring T.",
   title = "Navigation Along Windborne Plumes of Pheromone and Resource-Linked Odors", 
   journal= "Annual Review of Entomology",
   year = "2021",
   volume = "66",
   number = "Volume 66, 2021",
   pages = "317-336",
   doi = "https://doi.org/10.1146/annurev-ento-011019-024932",
   url = "https://www.annualreviews.org/content/journals/10.1146/annurev-ento-011019-024932",
   publisher = "Annual Reviews",
   issn = "1545-4487",
   type = "Journal Article",
  }

@article{Wechsler2023,
    author = {Wechsler, Samuel P. and Bhandawat, Vikas},
    title = {Behavioral algorithms and neural mechanisms underlying odor-modulated locomotion in insects},
    journal = {Journal of Experimental Biology},
    volume = {226},
    number = {1},
    pages = {jeb200261},
    year = {2023},
    month = {01},
    issn = {0022-0949},
    doi = {10.1242/jeb.200261},
    url = {https://doi.org/10.1242/jeb.200261},
    eprint = {https://journals.biologists.com/jeb/article-pdf/226/1/jeb200261/2436951/jeb200261.pdf},
}

@article{Louis2024,
title = {Drosophila flight: How flies control casts and surges},
journal = {Current Biology},
volume = {34},
number = {3},
pages = {R91-R94},
year = {2024},
issn = {0960-9822},
doi = {https://doi.org/10.1016/j.cub.2023.12.048},
url = {https://www.sciencedirect.com/science/article/pii/S0960982223017487},
author = {Matthieu Louis}
}

@article{Catania2013,
author = {Catania, Kenneth},
year = {2013},
month = {02},
pages = {1441},
title = {Stereo and serial sniffing guide navigation to an odour source in a mammal},
volume = {4},
journal = {Nature Communications},
doi = {10.1038/ncomms2444}
}

@article{Draft2018,
    author = {Draft, Ryan W. and McGill, Matthew R. and Kapoor, Vikrant and Murthy, Venkatesh N.},
    title = {Carpenter ants use diverse antennae sampling strategies to track odor trails},
    journal = {Journal of Experimental Biology},
    volume = {221},
    number = {22},
    pages = {jeb185124},
    year = {2018},
    month = {11},
    issn = {0022-0949},
    doi = {10.1242/jeb.185124},
    url = {https://doi.org/10.1242/jeb.185124},
    eprint = {https://journals.biologists.com/jeb/article-pdf/221/22/jeb185124/1905404/jeb185124.pdf},
}

@article{Jayakumar2022,
author = {Siddharth Jayakumar  and Venkatesh N. Murthy },
title = {A new angle on odor trail tracking},
journal = {Proceedings of the National Academy of Sciences},
volume = {119},
number = {3},
pages = {e2121332119},
year = {2022},
doi = {10.1073/pnas.2121332119},
URL = {https://www.pnas.org/doi/abs/10.1073/pnas.2121332119},
eprint = {https://www.pnas.org/doi/pdf/10.1073/pnas.2121332119}
}

@misc{rosa_github,
  author       = {},
  title        = {},
  howpublished = {\url{https://github.com/rosaflaquer/ChemotacticAnts}},
  year         = {}}

@article{Popp2023,
title = {Ants combine systematic meandering and correlated random walks when searching for unknown resources},
journal = {iScience},
volume = {26},
number = {2},
pages = {105916},
year = {2023},
issn = {2589-0042},
doi = {https://doi.org/10.1016/j.isci.2022.105916},
url = {https://www.sciencedirect.com/science/article/pii/S2589004222021897},
author = {Stefan Popp and Anna Dornhaus}
}

@article{vonThienen2014,
  author    = {Wolfhard von Thienen and Dirk Metzler and Dong-Hwan Choe and Volker Witte},
  title     = {Pheromone communication in ants: a detailed analysis of concentration-dependent decisions in three species},
  journal   = {Behavioral Ecology and Sociobiology},
  volume    = {68},
  number    = {10},
  pages     = {1611--1627},
  year      = {2014},
  publisher = {Springer},
  doi       = {10.1007/s00265-014-1770-3},
  url       = {https://link.springer.com/article/10.1007/s00265-014-1770-3}
}

@article{Lenoir2011,
  author    = {Alain Lenoir and Amélie Benoist and Abraham Hefetz and Wittko Francke and Xim Cerdá and Raphaël Boulay},
  title     = {Trail-following behaviour in two Aphaenogaster ants},
  journal   = {Chemoecology},
  volume    = {21},
  number    = {2},
  pages     = {83--88},
  year      = {2011},
  publisher = {Springer Basel},
  doi       = {10.1007/s00049-011-0071-9},
  url       = {https://link.springer.com/article/10.1007/s00049-011-0071-9}
}

@article{Cerda2009,
author={Cerd{\'a}, Xim
and Angulo, Elena
and Boulay, Rapha{\"e}l
and Lenoir, Alain},
title={Individual and collective foraging decisions: a field study of worker recruitment in the gypsy ant Aphaenogaster senilis},
journal={Behavioral Ecology and Sociobiology},
year={2009},
month={Feb},
day={01},
volume={63},
number={4},
pages={551-562},
issn={1432-0762},
doi={10.1007/s00265-008-0690-5},
url={https://doi.org/10.1007/s00265-008-0690-5}
}

\appendix

\newcounter{appendix_figs}
\setcounter{appendix_figs}{1}
\renewcommand\thefigure{\thesection.\arabic{appendix_figs}}   

\section{Experimental setup} \label{app:setup} \setcounter{appendix_figs}{1}

All experiments were conducted using one colony of approximatelly 150 \textit{Aphaenogaster Senilis} ants. The colony was collected on the Autonomous University of Barcelona campus in May, 2022 using standard protocols (digging up the nest and collecting the ants, including the queen, with the help of insect aspirators). Once collected, the ants were maintained in the laboratory in plastic boxes with a dark area connected to a water resource, and under controlled conditions until the end of the experiments.

The experimental setup consisted of a white plastic plate measuring $50 \times 23.5$ cm connected at its center to the ant nest (labeled as \textit{Nest} in Figure \ref{fig:ex_traj}a)). To prevent ants from leaving the plate, the entire structure was surrounded by water. In this setup, we painted a pheromone trail of $0.3$ cm wide in the shape of a continuous, oval loop: two parallel straight lines of $37$ cm joined by two circular regions of $6$ cm of radius (labeled as \textit{Trail} in Figure \ref{fig:ex_traj}a)).
The pheromone composition has been previously reported (see \cite{Lenoir2011}), being the alkaloyd anabaseine (3,4,5,6-tetrahydro-2,3$^{\prime}$-bipyridine) its main component. The pheromone was obtained from the ant's gasters following the protocol described in previous works \cite{vonThienen2014}.  Several individuals were kept in the freezer for 10-15 minutes, and later their gasters were dissected, and diluted in methanol (99,9\% pure) using a proportion of 0,5 ml/gaster. The gasters were gently milled with a steel mortar to extract the pheromone, and the dilution stored at 4ºC until use. This solution was then loaded into an empty alcohol-ink marker  of $0.3$ cm tip, which was used to deposit the pheromone trail on to the plate by drawing a continuous line. To ensure that only the introduced pheromone influenced behavior, the plate was covered with fresh parchment paper before each trial. A food patch consisting of mealworms cut into 2-3 pieces was placed one of the curved sections of the trail at the start of each run, to encourage exploration of the plate during successive experiments. 
The food source was chosen to be below the critical size and mass required for pheromone recruitment in \textit{Aphaenogaster senilis} \cite{Cerda2009}, ensuring that pheromone trails from previous ants are negligible and that all chemotactic interactions are in the \textit{Trail}.
The setup was then enclosed in an acrylic box and kept under uniform illumination and temperature.

Each experimental trial involved connecting the structure to the colony for one hour. The duration of the experiment was established to avoid the total evaporation of the chemical trail. We note that while the concentration decreases during the experimental run, the shape of the chemical trail remains constant. During this time, ant individuals were free to leave the nest, explore the plate and return to the nest. Each experiment was recorded with a video camera at $25$ Hz. Therefore, every single experiment, corresponding to one hour of recording, generated a total of $90,000$ frames. We repeated these trials over 20 days.

\section{Tracking}\label{app:tracking} 
\begin{figure}
    \centering
    \includegraphics[width=\linewidth]{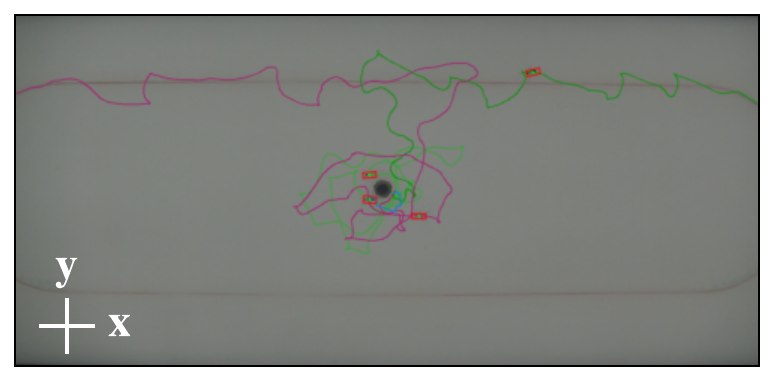}
    \caption{Example frame of the the trajectories detected during processing. All detected ants are boxed in red, with their mid point plotted in light green, which is the $(x,y)$ coordinates of our trails. In the plot we also present the full trajectories of the ants present in the frame.}
    \label{fig:Processing}
\end{figure}

\setcounter{appendix_figs}{1}
To obtain the ant's trajectories we used the software AnTracks \cite{AnTracks}. The workflow we used is as follows:
\begin{enumerate}
    \item Compute the averaged background image for the whole video, this is, the plate without ants.
    \item Pre-process each frame of the video by normalizing the image contrast to the full range of [0,225] (in black and white scale) and setting all the points below 100 contrast to the background.
    \item Use a \textit{Background Subtraction} method to find objects. All objects darker than the white background are identified.
    \item Split the objects of n times the average pixel size of the ants into n objects.
    \item Create the trajectory by trying to assign each object to its most likely trajectory of the previous frame.
\end{enumerate}
In AnTracks the likelihood that an object belongs to a particular trajectory is determined using a pair-cost function. In our case, we employed the \textit{Path-Prediction} approach: for each trajectory in the current frame, the next point is predicted via extrapolation of the previous particle movement. Object-trajectory pairing is chosen to minimize the distance between these predicted trajectory points and the centroids of the objects. A \textit{Global} optimization method performs a joint optimization for all object-trajectories assignments at the same time, and finds the globally best pairing. Since our video recordings are in a white plate and the ants are black, and that they are recorded in a controlled environment, this method is very robust. In Figure \ref{fig:Processing} we present an example frame of the ants detected and their respective trajectories.

\section{Data processing} \label{app:processing} \setcounter{appendix_figs}{1}

The trajectories of ants within the arena were extracted from video recordings using the AnTracks software \cite{AnTracks}. Ant velocities were then calculated as the discrete derivative of their positions. To identify interactions with the pheromone trail, a \textit{trail region} was defined as the area extending 1.14 cm above and below the trail (highlighted by the boxes in Figure \ref{fig:ex_traj}b). We have checked that small variations of this threshold definition do not qualitatively modify our results. 

Segments of trajectory where ants remained within 1.14 cm of the trail for more than 1.6 seconds were classified as \textit{near-trail} segments, while the remaining portions were labeled as \textit{not-in-trail}. Each \textit{near-trail} segment was then treated as an independent trajectory for further analysis. Next, to work with the data points when the ant is following the trail, we have truncated the last $t = 1.14/\left<v\right>$ seconds, corresponding to the ant escaping from the trail where $\left<v\right>$ is the average speed of each trajectory (in cm/s). Finally, from this set we worked only with trajectories of at least $150$ points (which corresponds to a trajectory duration of at least 6.0 seconds) to have sufficient points to compute meaningful statistics.
To standardize the dataset, all trajectories were rotated so that the trail aligned along the $(x,0)$ axis, regardless of whether the ants were located on the top or bottom side of the plate, and that $\theta \in [0, \pi/2]$ relative to the horizontal at $t=0$. From this processing we obtain a dataset of $157$ trajectories from which we disregard 1 as the ant, while being in the trail region it stopped moving. Finally, we have a dataset of $156$ trajectories.

\section{The overdamped limit} \label{app:overdamped} \setcounter{appendix_figs}{1}

In equation \eqref{eq:ds_ISM}, the force resulting from the chemotactic (DM) interaction appears in the spin derivative.
In other physical models, such as the continuous-time Vicsek model \cite{cavagna2024DLT}, the force acts directly on the velocity derivative. This is equivalent to assuming the overdamped limit in the ISM, where inertia is negligible compared to the dissipative forces. (\( \vert \chi \frac{d\vec{s}}{dt} \vert \ll \vert \eta\vec{s} \vert\)). Assuming this, we can isolate the spin $\vec{s}$ in \eqref{eq:ds_ISM} and substitute it in the velocity derivative \eqref{eq:dv_ISM}. Equation \eqref{eq:dv_ISM} now reads 
\begin{equation}
    \frac{d\vec{v}}{dt} = \frac{1}{\eta} \left(\vec{v} \times \left(-\frac{dH}{d\vec{v}}\right)\right) \times \vec{v}  +   \frac{1}{\eta} \vec{\xi} \times \vec{v}.
\end{equation}

Following the same procedure described in section \ref{sec:near}, and with $\vec{\xi}^*=(\vec{\xi} \times \vec{v})/\eta$, the equations for the evolution of the velocity components become

\begin{align}
    \frac{dv_x}{dt} &=  \frac{Dp}{\eta} v_y^2 +\xi_x^*\\
    \frac{dv_y}{dt} &=  -\frac{Dp}{\eta} v_x v_y + \xi_y^*.
\end{align}

By using that \( v_x = v_0 \cos \theta \), \( v_y = v_0 \sin \theta \), and doing an expansion for small $\theta$ we obtain 

\begin{equation}
    \frac{d \theta}{dt} = -\frac{Dpv_0}{\eta} \theta + \xi_x^*
\end{equation}

The temporal correlations of the angle $\theta$ are now different from the ones in \eqref{eq:corr}, and they read as
\begin{equation}
    \frac{\left< \theta(0) \theta(t) \right>}{\left< \theta(0)^2 \right>} = \exp\left(-\frac{Dpv_0}{\eta}t\right).
\end{equation}

This solution indicates that the velocity correlations $C_{v_y}(t)\approx C_\theta (t)$ should decay exponentially, without oscillations. Consequently, for an ISM-like description to reproduce the oscillatory behavior observed in the experiments, it must be underdamped. This implies that inertia is a relevant parameter and cannot be neglected.

\section{Goodness of the fit} \label{app:fit} \setcounter{appendix_figs}{1}

\begin{figure}[!ht]
    \centering
    \includegraphics[width=\linewidth]{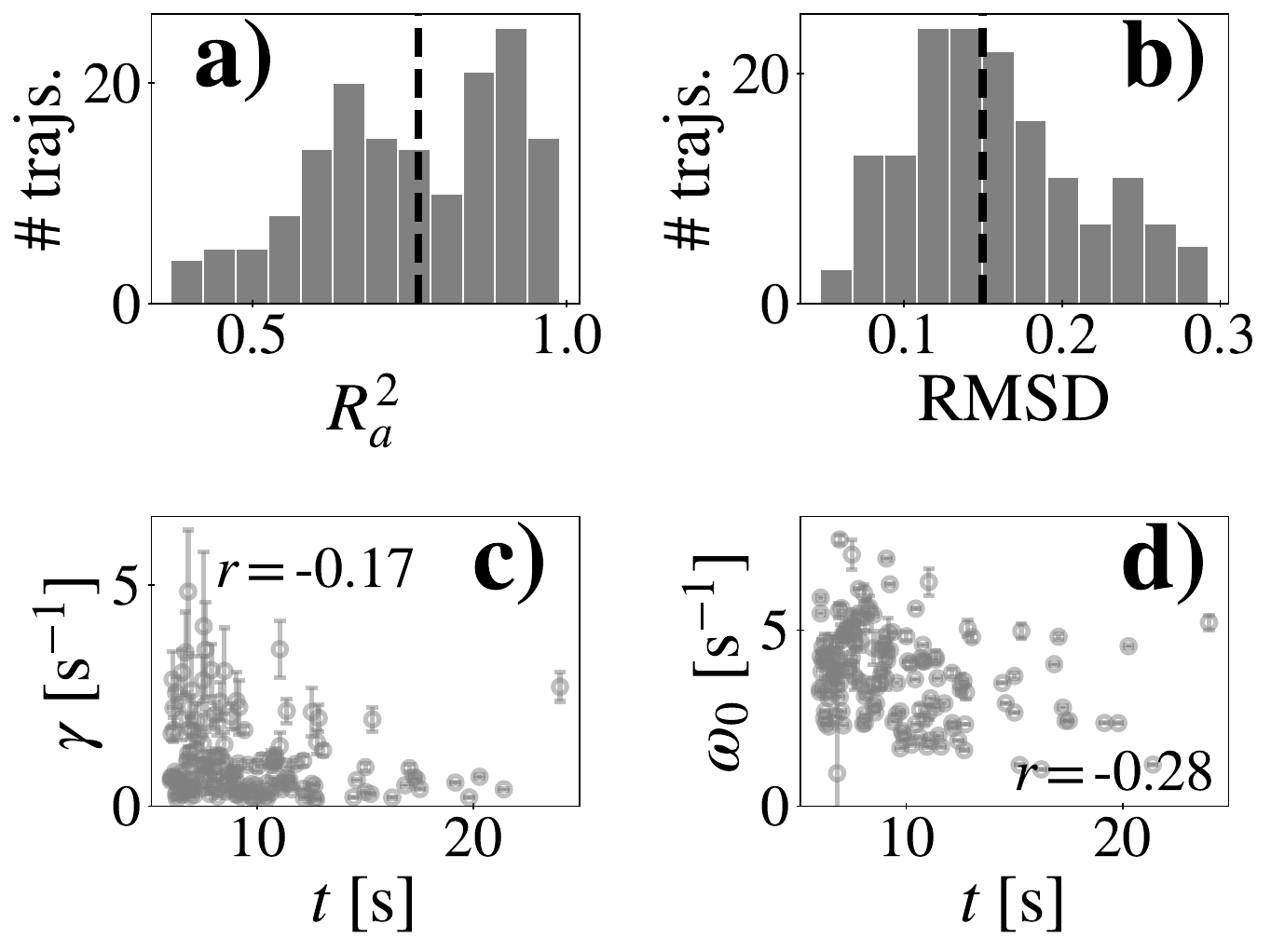}
    \caption{a) Histogram of the adjusted $R^2_a$. b) Histogram of the Root Mean Squared Error. In both cases, the dotted line represents the median of the distribution. b) Fitted damping parameter $\gamma$ as a function of the trajectory length $t$. c) Fitted frequency $\omega_0$ as a function of the trajectory length $t$. The value $r$ represents the correlation coefficient.}
    \label{fig:fit_metrics}
\end{figure}\addtocounter{appendix_figs}{1}

To fit Eq. \eqref{eq:corr} to the data points, we have used a non-linear least squares method, specifically the Levenberg-Marquardt algorithm. We analyze two goodness of fit metrics, the adjusted $R_a^2$ and the Root Mean Squared Error, defined as

\begin{align}
    R_a^2 &= 1-\frac{(1-R^2)(n-1)}{n-p-1} \nonumber \\
    RMSE &= \sqrt{\frac{1}{n}\sum_i^n(z_i-\hat{z}_i)^2}  \nonumber
\end{align}

where $R^2$ is the standard coefficient of determination, $n$ is the number of data points of each trajectory used in the fit, $p=2$ is the number of parameters to fit, $z_i$ are the data points, and $\hat{z}_i$ are the predicted ones. In Figure \ref{fig:fit_metrics} we present a histogram of the values obtained for the fits in Figure \ref{fig:triple} of the main text. For $R_a^2$ all values are higher than $0.4$ with a median of $R_a^2=0.76$. For the RMSD, all values are lower than $0.3$ with a median of RMSD$=0.15$. As Eq. \eqref{eq:corr} of the main text is bounded between $[-1,1]$ if values of RMSD$ \approx0.1$ represent $5\%$ of the range of the trajectory, which indicates that there are not large deviations between the correlations obtained from the data and the theory.

Furthermore, in Figure \ref{fig:fit_metrics} panels c) and d) we present the fitted parameters $\gamma$ and $\omega_0$ as a function of the trajectory duration. We compute their correlation coefficient, and in both cases, it yields a value of $r$ that indicates that the correlation is not very significant.

\section{$\omega_0$ vs $\left<v\right>$} \label{app:omega_vs_v}
\setcounter{appendix_figs}{1}

For completeness, we include the values of $\omega_0$ as a function of the characteristic mean speed $\left<v\right>$ in the same spirit as panels e), and f) in Figure \ref{fig:triple}.

\begin{figure}[!ht]
    \centering
    \includegraphics[width=0.5\linewidth]{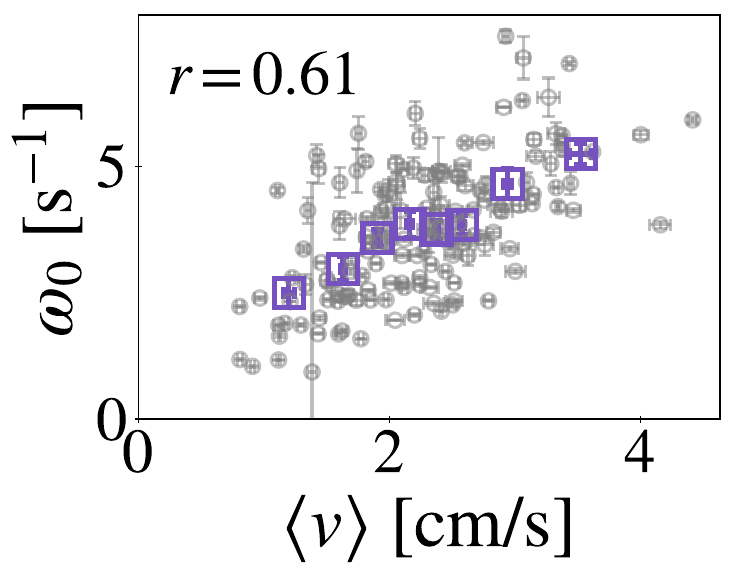}%
    \caption{Fitted values of $\omega_0$ as a function of the characteristic mean speed $\left<v\right>$ of each trajectory. The purple squared points correspond to an average of the $20$ points, grouping them according to their $\left< v \right>$ value. The value $r$ represents the correlation coefficient.}
    \label{fig:omega0_v}
\end{figure}\addtocounter{appendix_figs}{1}
\FloatBarrier

\end{document}